\documentclass{PoS}
\usepackage{bm,graphicx}

\PoS{PoS(LAT2005)298}

\newcommand{\naive}{\textsc{One-link }}

\newcommand{\hyp}{\textsc{Hyp }}
\newcommand{\hisq}{\textsc{Fat7xAsq }}

\def\slashchar#1{\setbox0=\hbox{$#1$}           % set a box for #1 
   \dimen0=\wd0                                 % and get its size
   \setbox1=\hbox{/} \dimen1=\wd1               % get size of /
   \ifdim\dimen0>\dimen1                        % #1 is bigger
      \rlap{\hbox to \dimen0{\hfil/\hfil}}      % so center / in box
      #1                                        % and print #1
   \else                                        % / is bigger
      \rlap{\hbox to \dimen1{\hfil$#1$\hfil}}   % so center #1
      /                                         % and print /
   \fi}                                         %

\newcommand{\be}{\begin{equation}}
\newcommand{\ee}{\end{equation}}
\newcommand{\bea}{\begin{eqnarray}}
\newcommand{\eea}{\end{eqnarray}}
\newcommand{\D}{\slashchar{D}}

\title{Index Theorem and Random Matrix Theory for Improved Staggered Quarks }

\ShortTitle{Index Theorem and Random Matrix Theory for Improved Staggered Quarks }

\author{\speaker{Eduardo Follana}\\
        University of Glasgow\\
        E-mail: \email{e.follana@physics.gla.ac.uk}}

\author{Alistair Hart\\
        University of Edinburgh\\
	E-mail: \email{a.hart@ed.ac.uk}}

\author{Christine T.H. Davies\\
        University of Glasgow\\
	E-mail: \email{c.davies@physics.gla.ac.uk}}

\abstract {We study various improved staggered quark Dirac operators
on quenched gluon backgrounds in lattice QCD. We find a clear
separation of the spectrum of eigenvalues into high chirality,
would-be zero modes and others, in accordance with the Index
Theorem. We find the expected clustering of the non-zero modes into
quartets as we approach the continuum limit. The predictions of random
matrix theory for the epsilon regime are well reproduced. We conclude
that improved staggered quarks near the continuum limit respond
correctly to QCD topology.}

\FullConference{XXIIIrd International Symposium on Lattice Field Theory\\
		 25-30 July 2005\\
		 Trinity College, Dublin, Ireland}

\begin{document}

\section{Motivation}

The low energy regime of Quantum Chromodynamics (QCD) exhibits a rich
and interesting phenomenology, including the $U_A(1)$ axial anomaly,
chiral symmetry breaking and the topological properties of the
theory. There are a number of detailed predictions of the properties
of these low-lying modes, such as the existence of an Index Theorem
\cite{Atiyah:1963,Atiyah:1968} and the distribution of the first few
eigenvalues in fixed topological charge sectors
\cite{Leutwyler:1992yt,Shuryak:1993pi,Damgaard:2001ep,Akemann:2003tv,Nishigaki:1998is,Damgaard:2000ah}.

Any correct discretization of QCD must agree with those predictions
close enough to the continuum limit. Here we show that this is already
the case for improved staggered quarks at the lattice spacings
typically used in present-day simulations. In particular, we show that
improved staggered fermions do respond correctly to the gluonic
topological charge, and discuss why some confusion on this issue
exists in the literature. Here we expand the work in
\cite{Follana:2004sz,Follana:2004wp}. 
For more detailed results see
\cite{Follana:2005prd}.
Related work has been presented in \cite{Wong1,Wong2}.

\section{Improved Staggered Dirac Operators}

The massless, gauge-invariant, \naive\ staggered Dirac operator on a
$d=4$ dimensional Euclidean lattice with spacing $a$ is
\be
\D (x,y)  =  \frac{1}{2au_0} \sum_{\mu=1}^d
\eta_\mu(x) \left[ U_\mu(x) \delta_{x+\hat{\mu},y} - H.c. \right] \; , \quad
\eta_\nu(x)  =  (-1)^{\sum_{\mu < \nu} x_\mu }
\label{eqn_naive}
\ee
with $u_0$ an optional tadpole-improvement factor given, in our case,
by the fourth root of the mean plaquette. $\D$ is antihermitian and
obeys a remnant of the continuum $\gamma_5$ anticommutation relation,
$
\left\{ \D,\epsilon \right\}  = 0
$
, with
$
\epsilon(x)  = (-1) ^ {\sum_{\mu=1}^d x_\mu} \;.
$
As in the continuum case, its spectrum is therefore purely imaginary,
with eigenvalues occurring in complex conjugate pairs, 
$
\{\pm i \lambda_s, \; \lambda_s \in \mathbb{R} \} \; .
\label{eqn_evalue_pairs}
$
The corresponding action describes $N_t = 4$ ``tastes'' of fermions
which interact via unphysical ``taste breaking'' interactions that
vanish in the continuum limit as $a^2$. In such limit there is an
$SU(N_t) \otimes SU(N_t)$ chiral symmetry, and the spectrum is
therefore $N_t$-fold degenerate. There is also an exact Index Theorem.
At finite lattice spacing the chiral symmetry group is reduced to
$U(1) \otimes U(1)$ and we do not expect to see this picture, for
example there will not be an exact Index Theorem anymore.

In addition to the \naive operator, we have also studied several
improved staggered operators, the so called \textsc{Asqtad}, \hisq and
\hyp, designed to suppress the taste-changing interactions
\cite{Naik:1986bn,Lepage:1998vj,Knechtli:2000ku}.

\section{Details of the Simulation}

We use a quenched, $SU(3)$ gluonic action that is both tree level
Symanzik and tadpole improved
\cite{Curci,Curcierratum,Luscher,Luscher_erratum,Alford}. It
differs from the one used by the MILC Collaboration
\cite{Davies:2003ik}
only in small one loop radiative corrections. We have generated
several ensembles of around 1000 configurations.

We measure the gluonic topological charge $Q$ using two different
cooling methods, to look for consistency. We discard the
configurations for which both methods disagree. In the $a=0.093$~fm
ensemble fewer than $10\%$ of the configurations have an ambiguous
topological charge.  For the finest ensemble, $a = 0.077$~fm, this
number goes down to around $2\%$. We stress that we only use cooling
to determine the topological charge. All the Dirac spectrum
measurements are done on the original thermalized (``hot'')
configurations.

\section{Index Theorem}
\begin{figure*}[t]
\includegraphics[width=6in]{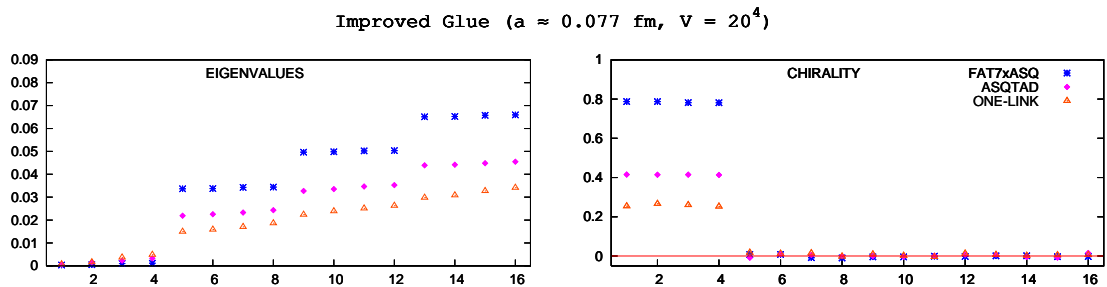}

\caption{\label{fig_spectrum} The positive half of a typical eigenmode
  spectrum for a configuration of $Q = 2$, in lattice units. The
  $x$-axis is eigenvalue number. The \hyp\ operator gives results very
  similar to \hisq, and they are not shown for clarity.}
\end{figure*}

We begin our analysis by qualitatively comparing the low-lying modes
of various staggered quark operators on a typical gauge background
selected from those with $Q~=~2$. In Fig.~\ref{fig_spectrum} we show
both the eigenvalue and the chirality of the first sixteen (positive)
eigenmodes. Near the continuum limit we expect to see first $2|Q|=4$
near-zero modes with their chirality renormalised slightly away from
unity. The remaining modes should have chirality near zero and divide
into almost degenerate quartets. The spectrum looks quite
continuum-like for all Dirac operators and we see a clear Index
Theorem. For the improved operators we also see a very clear quadruple
degeneracy in the non-zero modes. The renormalisation of the chirality
away from 1 is small for the improved Dirac operators (around
$Z=1.2$), becoming as large as $Z \approx 4$ for the \naive\
fermions. This result is fairly typical of operator renormalisation
for staggered quarks \cite{Hein,Trottier:2002,Lee:2002}.

\section{Comparison with Random Matrix Theory}

The chiral symmetries of staggered quarks are more complicated than
for continuum QCD. At finite lattice spacing there are $N_t^2=16$ pion
states, and only one of these becomes massless in the chiral limit,
the ``Goldstone pion'', with mass $M_G$.  The 15 remaining states
have masses $\sim M_{NG}$ arising from the taste breaking
interactions. The (chirally extrapolated) masses are
$\mathcal{O}(a^2)$ and therefore zero only in the continuum limit.

There are potentially \textit{two} universal regions for different
parameters of the system:
\begin{eqnarray}
\mbox{$\varepsilon$-regime:} ~~~ &&
(\Lambda_{QCD})^{-1} \ll L \ll (M_{NG})^{-1} \; ,
 \\
\mbox{$\varepsilon^\prime$-regime:} ~~~ &&
(M_{NG})^{-1} \ll L \ll (M_{G})^{-1} \; .
\end{eqnarray}
In the first one the chiral symmetries are as in the continuum,
corresponding to an effective theory of 16 massless pions. The
low-lying non-zero modes have a near $N_t$-fold degeneracy, and follow
the distributions corresponding to continuum QCD (and chiral lattice
fermions
\cite{Edwards:1999ra,Edwards:1999zm,Damgaard:1999tk,
Hasenfratz:2002rp,Bietenholz:2003mi,Giusti:2003gf}).

At finite lattice spacing there is a second,
$\varepsilon^\prime$-regime, corresponding to an effective theory with
a single massless pion. There is not even approximate restoration of
the continuum symmetries, and the associated RMT has only $U(1)
\otimes U(1)$ chiral symmetry.  The universal predictions for the
eigenvalues are therefore strikingly different from those for
continuum QCD. In particular, the predictions are the same for all
sectors of topological charge
\cite{Damgaard:PrivComm}.
Presumably it is this regime that was studied in
\cite{Berbenni-Bitsch:1998tx,Damgaard:1998ie,
  Gockeler:1998jj,Damgaard:1999bq,Damgaard:2000qt}.
With the coarse lattices and unimproved gauge ensembles used in these
studies, they observed no sensitivity of the eigenvalue spectra to
$Q$, leading to the incorrect folklore that staggered quarks are
``blind to the topology''.
To see the continuum chiral symmetries, there must be a sufficiently
large mass gap between the heaviest pion and the lightest of the other
hadrons. This requires the use of improved gauge and fermion actions
and a sufficiently small lattice spacing. At coarse lattice spacings,
with unimproved fermions, $M_{NG} \approx \Lambda_{QCD}$
and there will be no $\varepsilon$-regime.

\begin{figure*}[t]
\includegraphics[width=5.in,height=5.3in,clip]{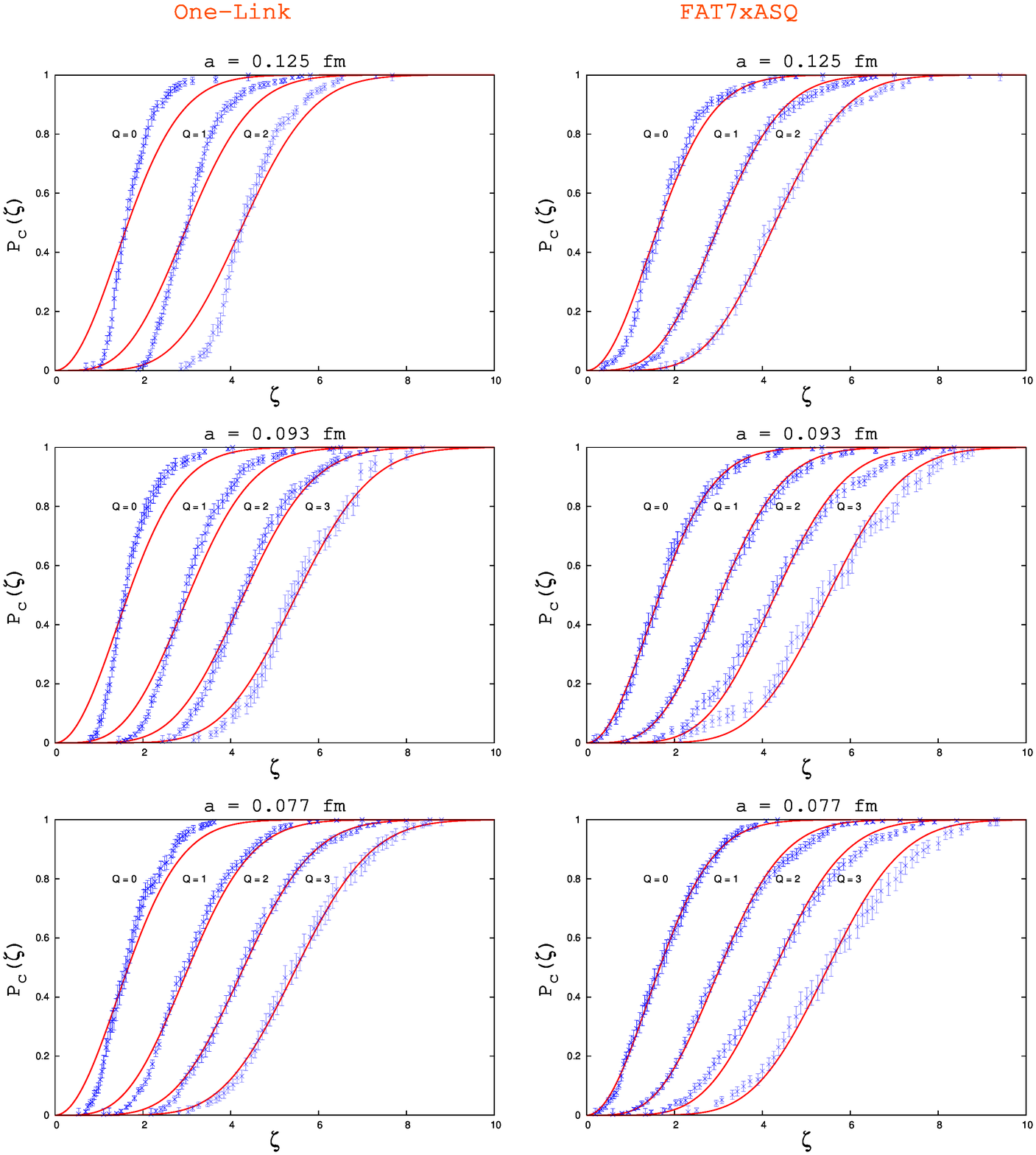}

\caption{\label{fig_rmt_fits_k1} Comparison of the unfolded eigenvalue
  distribution with RMT for the lowest eigenvalue for different
  topological charges. The lattice size is kept at $aL \approx 1.5$~fm}
\end{figure*}

We separately fit the individual cumulative spectral densities to the
predictions from RMT. These one parameter fits yield a prediction for
the chiral condensate.
We show some examples of the fits in Figs.~\ref{fig_rmt_fits_k1}.
It is clear that the improved Dirac operators match the predictions of
RMT very closely over the full range of lattice spacings. By contrast,
even at $a=0.077$~fm the \mbox{\naive} Dirac operator still shows
significant discrepancies. The fits also show clearly the expected
variation with topological charge. Each fit gives a value for the bare
chiral condensate, and we find all of them to be broadly consistent
with each other.

\section{Conclusions and Outlook}

We have studied the low-lying eigenmodes for the staggered lattice
Dirac operator, using a range of improved and unimproved versions. The
improved operators pass all the tests. In particular, and contrary to
previous accepted wisdom, such fermions do respond exactly as expected
to the gluonic topological charge, and in a way identical to continuum
QCD and other lattice formulations of the Dirac operator.

We have seen that for improved operators the eigenvalue spectrum
divides cleanly into near-zero and non-zero modes. The near-zero modes
are characterised by a uniformly high chirality, with their relative
number fixed by the Atiyah-Singer Index Theorem. The non-zero modes,
by contrast, have chirality that is near zero and they divide into
near degenerate quartets. In addition, the low-lying non-zero modes
follow closely the universal distributions predicted by random matrix
theory.

\begin{acknowledgments}
  
This research is part of the EU integrated infrastructure initiative
hadron physics project under contract number RII3-CT-2004-506078.
  
We thank: Ph.~de~Forcrand for his topological charge measurement code;
A.~Hasenfratz for help in implementing the \hyp\ operator;
G.P.~Lepage, P.~Damgaard and G.~Akemann for useful discussions. E.F.,
C.T.H.D. and Q.M. are supported by PPARC and A.H. by the U.K.  Royal
Society.  The eigenvalue calculations were carried out on computer
clusters at Scotgrid and the Dallas Southern Methodist University. We
thank David Martin and Kent Hornbostel for assistance.

\end{acknowledgments}

\bibliographystyle{h-physrev4}
\bibliography{proceeding}

%\begin{thebibliography}{99}
%  \bibitem{...} ....
%\end{thebibliography}

\end{document}